\documentstyle[preprint,aps]{revtex}
\begin{document}
\preprint{Running head: Josephson flux-flow oscillators}
\draft
\title{Coupling of Josephson flux-flow oscillators\\ to an external RC load}
\author{C. Soriano, G. Costabile, R. D. Parmentier\cite{byline1}}
\address{Research Unit INFM and Department of Physics,\\
University of Salerno,\\
I-84081 Baronissi (SA), Italy}
\date{2 April 1996}
\maketitle

\begin{abstract}
We investigate by numerical simulations the behavior of the power 
dissipated in a resistive load capacitively coupled to a Josephson flux 
flow oscillator and compare the results to those obtained for a d.c. 
coupled purely resistive load. Assuming realistic values for the parameters 
$R$ and $C$, both in the high- and in the low-$T_c$ case the power is large
enough to allow the operation of such a device in applications.
\end{abstract}


\section{Introduction}

Low noise measurements in the sub-millimeter range, {\it e.g.}, in 
radioastronomy, require a stable, low noise local oscillator for the 
receiver. The Flux Flow Oscillator (FFO), a device made with a long 
Josephson junction, has been proposed as a good candidate for this 
application \cite{Koshelets95}. Some of the most important features of such 
a device are the following \cite{Nagatsuma84,Kosh93}: the output power 
is relatively large ($0.1 - 1\, \mu W$), the oscillator can be easily tuned 
in a wide band ($75 - 500\, GHz$) by varying the bias and a magnetic field,
and the emitted signal has a very narrow linewidth ($130 \,kHz$ at 
$70\,GHz$ \cite{Ustinov92}, less than $2.1\, MHz$ in the band $280 - 330\, 
GHz$ \cite{Zhang93a}). As the signal generated by a local oscillator has to 
be coupled to a mixer or to a transmission line, in the literature 
different couplings to a load have been realized and studied 
\cite{Kosh93,Zhang93a,Mat95,Cirillo91,Nagatsuma83,Nagatsuma85,Zhang93b}. In
some works, in particular, the FFO has been d.c. coupled to a small 
junction acting as a detector, or, possibly, as a mixer; in this case, the 
real drawback is that the junctions cannot be biased independently. 
Capacitive coupling not only overcomes this drawback, but it also allows to 
increase the power transferred to the load, since it eliminates the d.c. 
loading of the oscillator. In this paper we investigate and compare these 
two coupling techniques for an FFO. We note in passing that ours is not the
first work to propose the use of a d.c. block between oscillator and load
device--see,{\it e.g.}, \cite{Kosh93}. However, in contrast with
\cite{Kosh93}, which described a complete and detailed integrated subsystem,
we focus specifically on the effects of the coupling element on system
performance.

\section{The physical device - the flux flow oscillator}

The FFO consists of a long Josephson junction biased by a d.c. current 
$I_B$ and driven by the effect of a magnetic field $H_e$, perpendicular to 
the length $L$ of the junction and parallel to the barrier, into a 
dynamical state in which the unidirectional motion of flux quanta 
(Josephson current vortices, or {\it fluxons}) takes place 
\cite{Parmentier93}. With reference to the configuration sketched in 
Fig.~\ref{layout}, the fluxons continuously penetrate from the left edge of 
the junction and propagate to the right; this regime corresponds to the 
appearance of typical branches in the $I-V$ characteristic. An array of 
vortices travels in the junction with phase velocity 
$u=(V_{dc}/\overline{c}d\mu_{0}H_{e})\overline{c}$, where $V_{dc}$ is the 
average voltage across the junction, $H_e$ is the external magnetic field 
applied in the $-y$ direction, $\overline{c}$ is the propagation velocity 
of electromagnetic waves in the junction and $d$ is the effective magnetic 
thickness of the barrier. The largest signal in the flux flow regime is 
obtained biasing the junction at the top of the flux flow branch in the   
$I-V$ characteristic; the corresponding voltage is determined by the {\it 
velocity matching condition\/}: the velocity $u$ approaches the Swihart 
velocity $\overline{c}$ and, hence, $V_{dc}\cong \left( 
\overline{c}d\mu_{0} \right) H_{e}$. One should take into account that a 
possible self magnetic field and the focusing effect of the external 
magnetic field \cite{Nagatsuma84} could make $V_{dc}$ larger. We shall
assume that in the stationary state the expression for the junction voltage 
is given by the Fourier series \cite{Irie85}

\begin{equation} %% 1
v(x,t)=V_{dc}+\sum_n^{\prime}V_n(x)\exp (jn\omega_{FF}t) \, ,
\label{series}
\end{equation}
where the coefficients $V_n$ are complex amplitudes and the 
$\sum_n^{\prime}$ denotes summation from $n=-\infty$ to $n=+\infty$, but 
without the term $n=0$, and $\omega_{FF}=(2\pi /\Phi_{0})V_{dc}$ is the 
fundamental angular frequency of the series. Since the amplitude of the 
fundamental component of the series~(\ref{series}) is dominant, the 
frequency of the FFO signal is

\begin{equation} %% 2
f_{FF}=V_{dc}/\Phi_{0} \, .
\label{frequency}
\end{equation}
This is just the frequency expected from the Josephson relations when a 
d.c. voltage $V_{dc}$ is applied across a junction.

An important parameter for an oscillator is the frequency range in which 
it can generate a signal useful for applications, or, in other terms,
its effective  bandwidth. Since, under the velocity matching 
condition $(u\rightarrow \overline{c})$, the frequency of the FFO is 
proportional to the magnetic field, 
$f_{FF}=(\overline{c}d\mu_{0}/\Phi_{0})H_{e}$, the effective bandwidth of 
the FFO is bounded to the values of the magnetic field within which the 
flux flow dynamic state is stable. Its lower limit is given by the minimum 
value of magnetic field needed to have penetration of fluxons into the 
junction, {\it i.e.\/}, by the critical field
$H_{e\min}=2j_{c}\lambda_{J}$. The upper limit can be estimated by 
supposing the maximum voltage in the flux flow regime to be of the order of 
the voltage gap (or of the order of $R_{n}I_{c}$ in the high-$T_c$ context),
as the nonlinear internal dynamics is strongly attenuated above that value. 
The tuning of the FFO to the desired frequency is achieved by changing
the magnetic field to move the flux flow branch and by setting the d.c. 
current to bias the junction near the top of the branch.

\section{Model and computational techniques}

In Fig.~\ref{layout} there is a schematic representation of a long 
Josephson junction, the FFO, that produces a signal coupled to the load 
through a capacitance. For this investigation the load device is assumed to
be purely resistive. This assumption is made with the hypothesis that such a
load model is sufficiently accurate to give information on the practical 
advantages of the capacitive coupling in comparison with simple d.c.
coupling; for other loads, an appropriate model could be chosen and
analyzed using the same approach.

The mathematical model describing the flux-flow oscillator is the perturbed 
sine-Gordon equation (PSGE), which in normalized form is

\begin{equation} %% 3
\varphi _{xx}-\varphi _{tt}-\sin \varphi =\alpha \varphi _t-\beta 
\varphi_{xxt}-\gamma \, ,
\label{PSGE}
\end{equation}
with the boundary conditions

\begin{mathletters}
\begin{equation} %% 4a
\varphi_x(0,t)+\beta \varphi_{xt}(0,t)=-\eta \, ,
\label{bc(0)}
\end{equation}
\begin{equation} %%4b
\varphi_x(L,t)+\beta \varphi_{xt}(L,t)=-\eta -i_{L}(t) \, .
\label{bc(L)}
\end{equation}
\end{mathletters}
Here, $\varphi$ is the phase difference between the junction electrodes, 
$x$ is the spatial coordinate normalized to the Josephson penetration 
length $\lambda_{J}$, $t$ is the time normalized to the inverse of the 
Josephson angular plasma frequency $\omega_{J}$, the subscripts indicate 
partial differentiation, the term in $\alpha$ represents shunt loss due to 
quasiparticle tunneling (here assumed ohmic), the term in $\beta$ represents
surface loss in the junction electrodes, $\gamma$ is the distributed bias
current $j_{B}$ normalized to the critical current density $j_{c}$, $L$ is
the normalized junction length, $\eta$ is the external magnetic field in the
plane of junction and perpendicular to its long dimension in units
normalized to $j_{c}\lambda_{J}$. In the $R-C$ load network $i_{L}$ is the
current normalized to $j_{c}\lambda_{J}w$ (where $w$ is the width of the
junction), $R_{L}$ is the load resistance normalized to the (linear)
characteristic impedance of the junction $Z_{0}$, $C_{L}$ is the coupling
capacitance normalized to the capacitance $C_{0}=1/\omega_{J}Z_{0}$ and 
$\omega_{L}=1/R_{L}C_{L}$ is the load angular frequency normalized to 
$\omega_{J}$; accordingly,the equation for the current $i_{L}$ is

\begin{equation} %% 5
\frac{di_{L}}{dt}=-\omega_{L}i_{L}+\frac{\varphi_{tt}(L,t)}{R_{L}} \, .
\label{iL}
\end{equation}

\section{Numerical results}

Eqs.~(\ref{PSGE},~\ref{iL}) with the boundary conditions
Eqs.~(\ref{bc(0)},~\ref{bc(L)}) have been integrated numerically using a 4th
order Runge-Kutta algorithm on a spatially discretized counterpart, varying
$R_{L}$ from 0.1 to $10^3$ and $C_{L}$ from $10^{-2}$ to 10. In the 
following we list the most relevant results. We note in passing that
Fig.~\ref{layout} depicts schematically a sandwich-type tunnel junction
structure. Although the fabrication technology for high-$T_c$ tunnel
junctions is not yet completely mature, significant progress in this
direction is being made \cite{highTc}. Accordingly, in what follows we have 
chosen model parameter values that are aimed in the direction of describing 
a lightly-hysteretic tunnel junction.

In Fig.~\ref{FFT} we report the Fourier spectrum of the voltage at the
right edge of the junction $v(L,t)$ calculated over more than two hundred 
periods. The spectrum consists of a d.c. component, whose height is the 
average voltage $V_{dc}$, a fundamental line at the frequency $f_{FF}$, 
having amplitude $V_{ac}(f_{FF})$, and a number of harmonics, up to the 
7th; the small background is due to the finite sampling, to the finite 
integration time and to numerical noise. We remark that the fundamental 
frequency turns out to be just equal to what one would calculate from the 
height of the d.c. component (taking into account the normalization), in 
full agreement with Eq.~(\ref{frequency}). Moreover, one sees that the 
fundamental harmonic is dominant and the amplitude of the other harmonics 
decreases exponentially with increasing order, as can also be inferred 
qualitatively from the inset in the same figure. We did not calculate the 
voltage spectrum for every value of the load used in our simulations; 
rather, we performed for most of them a simpler test to check that the 
series~(\ref{series}) could be truncated to the second term. In fact, we 
calculated the rms amplitude of the fundamental harmonic 
$V_{ac}(f_{FF})/\sqrt{2}$ from the signal $v(L,t)$ using the formula for 
the Fourier series coefficient and the rms value of the a.c. component of 
the signal $v_{ac}(L,t)\equiv v(L,t)-V$; the comparison shows that they are 
always equal within 1\%, so that in practice all the power is in the 
fundamental harmonic. In other words, in the case of the $R-C$ load the 
flux flow signal $v_{ac}(L,t)$ maintains, with good approximation, a 
sinusoidal form for every value examined of $R_{L}$ and $C_{L}$. Therefore, 
the output power $P_{L}$ is assumed to be the power of the first harmonic 
of the signal transferred from the flux flow oscillator to the resistive 
load $R_{L}$. For our $R-C$ loading network it is given by

\begin{equation} %% 6
P_{L} = \frac{V_{ac}^2(f_{FF})}{2R_{L}\left[ 1+\left( \omega 
_{L}/\omega_{FF}\right)^2\right]} \, .
\label{PL}
\end{equation}
Here $P_{L}$ is normalized to the Josephson power $P_{u}=V_{J}^{2}/Z_{0}$, 
where $V_{J}=\Phi_{0}\omega_{J}/2\pi$ is the normalizing Josephson voltage.

We compared the output power for the $R-C$ load to that of d.c. coupling to 
a resistive load, which has been already studied by Zhang \cite{Zhang93c}, 
to emphasize the different behavior. We have, in our model, the pure-$R$ 
load case by setting $\omega_{L}=0$ in Eq.~(\ref{iL}) and Eq.~(\ref{PL}). 
For the sake of consistency, we used the same parameters of Ref.~ 
\cite{Zhang93c}, {\it i.e.}, $\alpha=0.25$, $\beta=0.005$, $L=20$, biasing 
the junction with $\gamma =1.25$ on a flux flow step obtained with 
$\eta=4$; the results essentially agree with those in \cite{Zhang93c}, 
differences being attributable to the fact that with purely resistive 
coupling the output waveform is no longer a clean sinusoid, so the total 
power differs from the first-harmonic power. The maximum output power 
dissipated by the a.c. component is $P_{Lmax}=0.76$ for $R_L=7$, and 
changes very little in the range 3--10 (we note parenthetically that these 
numbers are reminiscent of those obtained in an early study \cite{Erne81} 
of a resonant-fluxon oscillator with pure-$R$ loading). This value is
represented in Fig.~\ref{PvsR} and Fig.~\ref{PvsC} as a horizontal straight
line for the sake of comparison with the $R-C$ case.

The $R-C$ case is, at first sight, more complex, as has already been noted 
by Zhang (see Section 4.4 of \cite{Zhang93c}). In Fig.~\ref{PvsR} we report
the output power as a function of the resistance $R_{L}$ for fixed values of
the ratio $\omega_{FF}/\omega_{L}$. We see that the power increases as the
oscillator frequency grows with respect to the load characteristic
frequency, and that it is not a monotonic function of the resistance. In
fact, maximum output power is obtained for $R_{L} \simeq 1$, {\it i.e.},
when the load resistance $R_{L}$ is close to the junction characteristic
impedance $Z_{0}$. This result is consistent with what one should expect
from considering the best matching condition in the framework of (linear) 
microwave transmission line theory. We remark also that the variation
of the output power is negligible for $\omega_{FF}/\omega_{L} \geq 5$. In 
Fig.~\ref{PvsR} the comparison with the curve with $P_{L}=0.76$ shows in 
which range the $R-C$ loaded FFO is more efficient than the $R$ loaded FFO.

In Fig.~\ref{PvsC}, in order to provide a more straightforward tool for 
device design, we plot the output power as a function of the load 
capacitance for fixed values of the load resistance $R_{L}$. For a given 
value of $R_{L}$, the power is first an increasing function of $C_L$ which, 
however, quickly saturates. The asymptotic values lie on the curve with 
$\omega_{FF}/\omega_{L}=100$ of Fig.~\ref{PvsR}, which well approximates 
the case of infinite capacitance.

Whereas, as mentioned above, the $R-C$ case appears at first to be more 
complicated than the pure-$R$ case, in fact it is simpler: once the values
of $V_{ac}$ and $\omega_{FF}$ in Eq.~\ref{PL} are established by the
junction dynamics, at least for the parameter values that we have studied,
the oscillator behaves with respect to the load much as a simple linear 
oscillator, characterized by a fixed internal voltage and a fixed internal 
resistance equal to $Z_{0}$. This operational simplicity is obtained 
essentially from the elimination of the d.c. loading of the oscillator, 
which, instead, is present in the pure-$R$ case.

\section{Discussion}

As we have seen in the previous section, in order to maximize the output 
power for a given value of the resistance, it is recommendable to increase 
the load capacitance, because the best coupling is obtained for $C_{L} \geq 
1$. To check whether this indication can be translated into realistic 
parameters to design and fabricate a device, we shall estimate what one 
should expect in two practical cases. In both cases we shall estimate 
$C_{p}$ from the formula $C_{p}=\epsilon_{r}\epsilon_{0}S/d$, with the 
usual meaning of the parameters. First, we shall consider the high-$T_c$ 
case, and we shall suppose that the coupling capacitor is a typical silicon 
monoxide element ($\epsilon_{r} \cong 5$, $d=600$ nm and $S=50 \mu$m 
$\times 50 \mu$m). Since $\omega_{J}=\sqrt{2\pi j_{c}/(\Phi_{0}C)}$ and 
$Z_{0}=(1/w)\sqrt{\mu_{0}d/C}$ where $C$ is the junction capacitance per 
unit area, the previous definitions give:

\begin{equation} %% 7
C_{L}=\frac{C_{P}}{wC}\sqrt{\frac{2\pi \mu_{0}j_{c}d}{\Phi_{0}}} \, .
\label{CL}
\end{equation}
Assuming from the literature $j_{c}=20$ kA/cm$^2$, $w=200$ nm, $d=280$ nm, 
$C=30$ fF/$\mu$m$^2$, we finally get $C_{L}=15.7$. To evaluate the power 
dissipated in the load, we can use the same figures, and find that 
$P_{u}=0.13\,\mu$W; this gives an idea of the order of magnitude of the 
power that could be extracted.

Next, let us consider a low-$T_c$ oscillator. In this case, the coupling 
capacitor can be made by growing a niobium oxide film by anodization 
($\epsilon_{r} \cong 25$, $d=100$ nm). From the previous formula, assuming 
typical parameters for Nb-AlO$_{x}$-Nb Josephson junctions ($j_{c}=1$ 
kA/cm$^2$, $w=3\,\mu$m, $d=80$ nm, $C=60$ fF/$\mu m^2$), we find 
$C_{L}=2.3$. Of course, in this case one is left with the problem of 
fabricating a load resistor that should be perhaps 0.1 ohm (or less), but
this is feasible with present-day technology; moreover, our numerical 
simulations should be considered as being merely indicative, in that the 
dissipation in a low-$T_c$ junction can be significantly lower than the 
value considered in this paper, which might bring about qualitative changes 
in the dynamics. Nevertheless, we think that it is interesting to take into
account a design based on the well established niobium technology, and we
shall explore this topic in more detail in the future.

\acknowledgements

We are grateful to J.\ E.\ Nordman for a critical reading of the manuscript
and for several useful suggestions. This work was supported by EU Contract
ESPRIT HTSC-GBJ 7100.

\begin{figure}
\caption{Schematic layout of flux-flow oscillator capacitively coupled to a 
load.}
\label{layout}
\end{figure}

\begin{figure}
\caption{Fourier spectrum of the output voltage. The waveform is shown in 
the inset. The parameters used in this calculation are: $R_L=1.5$, 
$C_L=0.01$, $V_{dc}=4.218$, $\eta=4$, $\alpha=0.25$, $\beta=0.005$, 
$\gamma=1.25$, $L=20$. The amplitude is normalized to $V_J=\Phi _0\omega 
_J/2\pi$; the frequency is normalized to $f_{FF}=0.671$.}
\label{FFT}
\end{figure}

\begin{figure}
\caption{Output power as a function of load resistance for different values 
of the ratio between the FFO frequency and the $R-C$ characteristic 
frequency. Here $\eta=4$, $\alpha=0.25$, $\beta=0.005$, $\gamma =1.25$, 
$L=20$.}
\label{PvsR}
\end{figure}

\begin{figure}
\caption{Output power as a function of load capacitance for different 
values of the load resistance. Parameters are as in Fig. 3.}
\label{PvsC}
\end{figure}

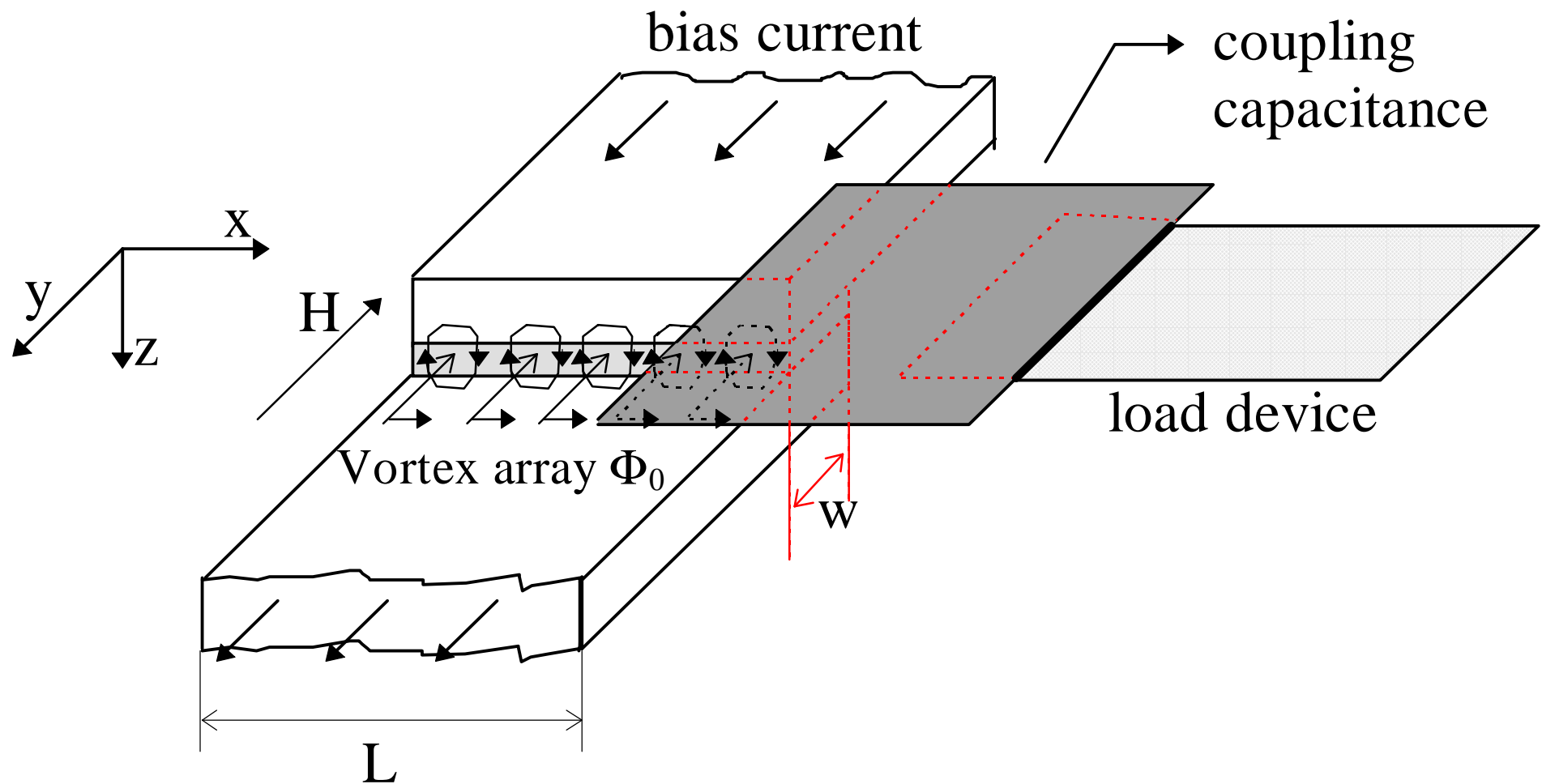
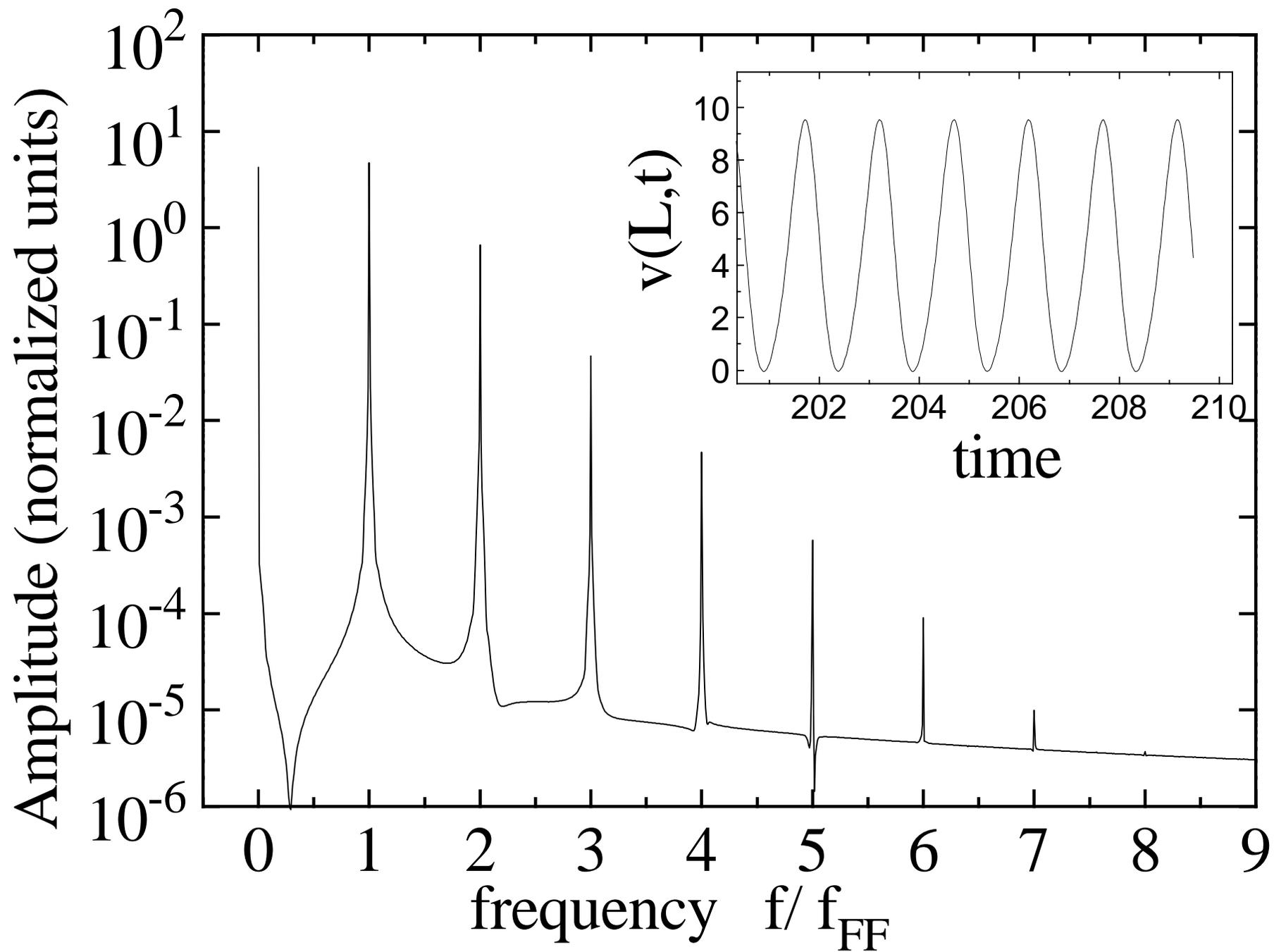
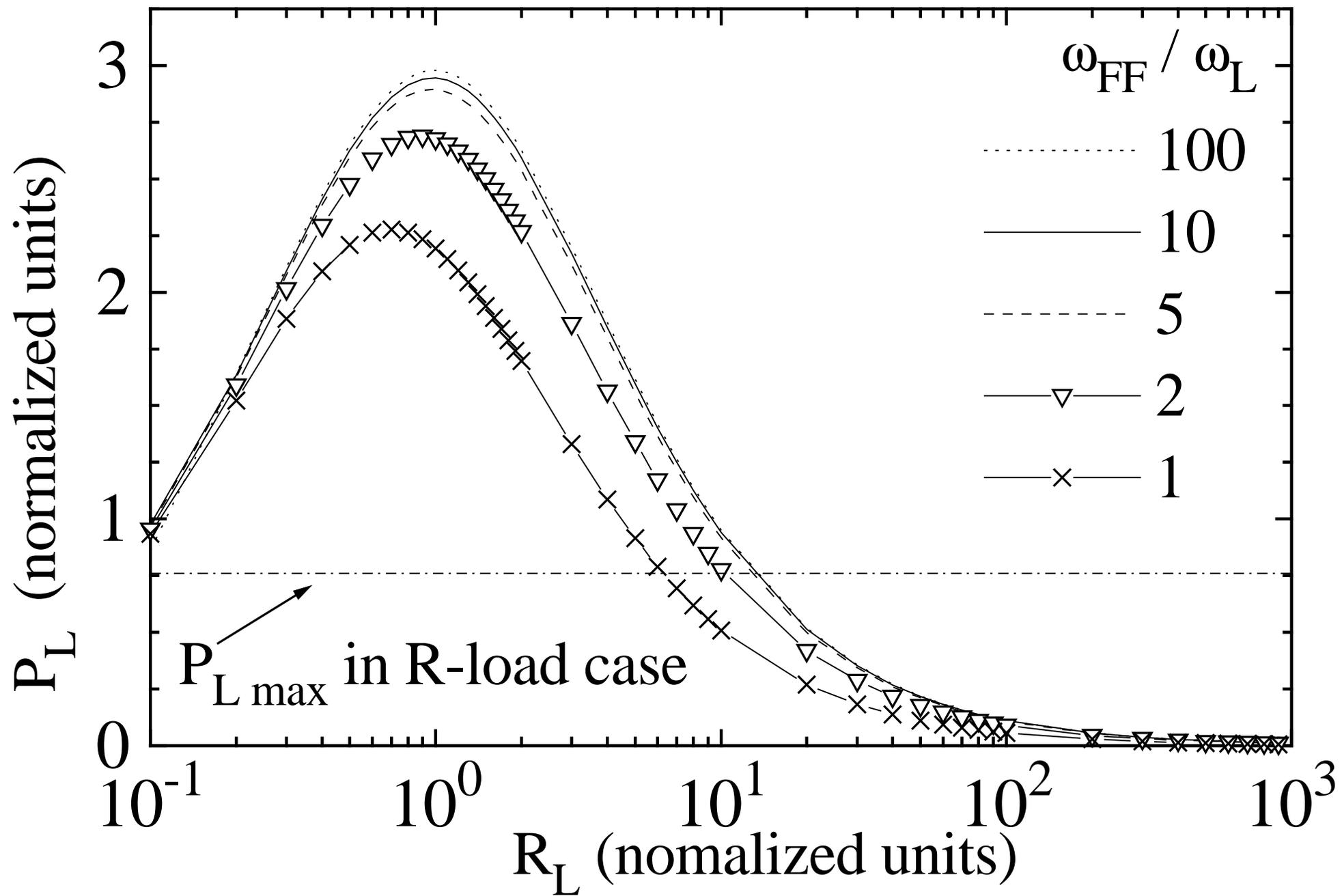
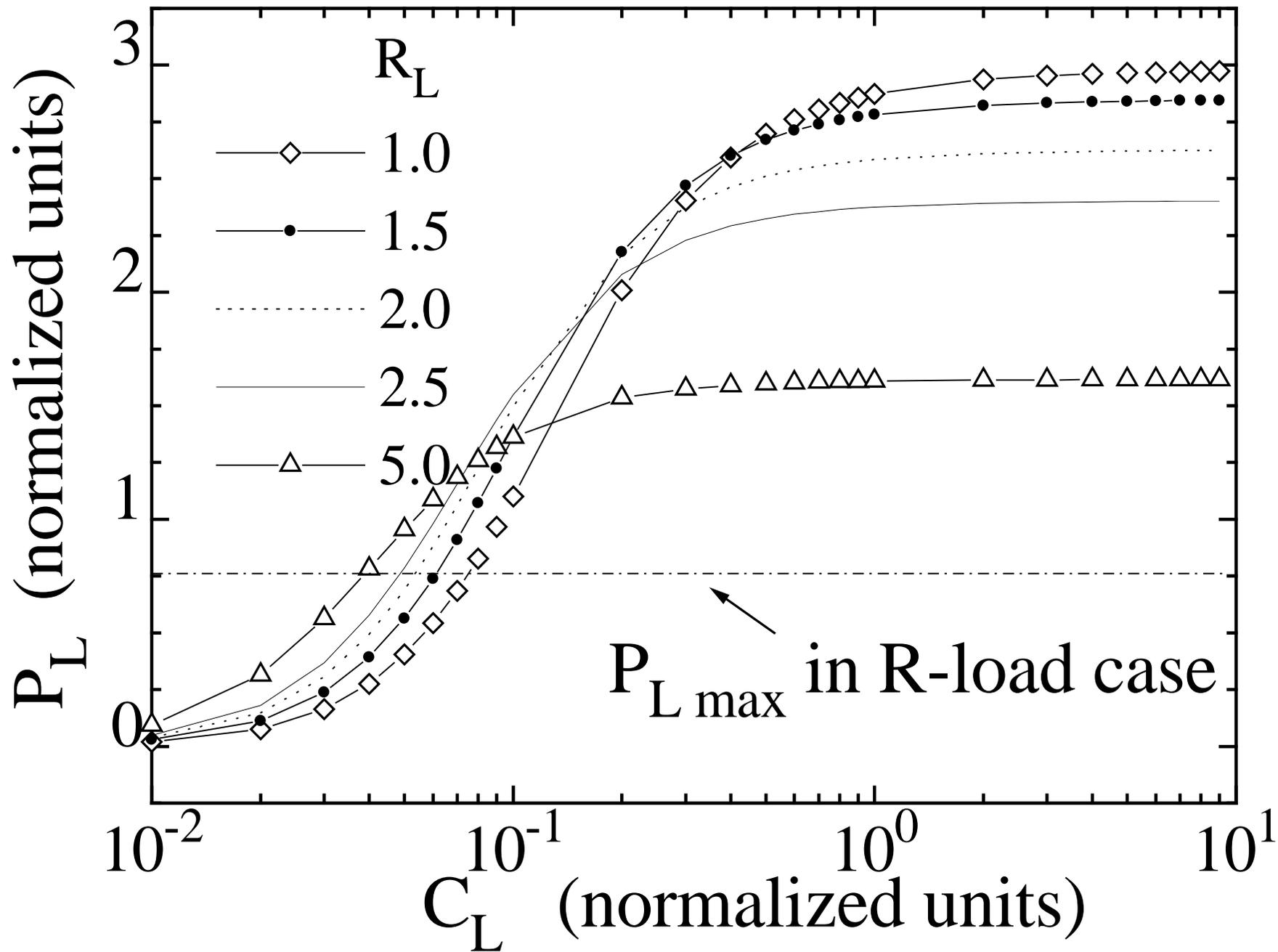
\end{document}